\documentclass[twocolumn,aps,a4paper,prl,numerical,superscriptaddress]{revtex4-1}

\usepackage{amsmath}
\usepackage{epsf}
\usepackage{graphicx}
\usepackage{natbib}
\usepackage{times}

\begin{document}

\title{Time-of-Flight Measurements of Single-Electron Wave Packets in Quantum-Hall Edge States}

\author{M.~Kataoka}
\email{E-mail: masaya.kataoka@npl.co.uk }
\affiliation{National Physical Laboratory, Hampton Road, Teddington, Middlesex TW11 0LW, United Kingdom }
\author{N.~Johnson}
\affiliation{National Physical Laboratory, Hampton Road, Teddington, Middlesex TW11 0LW, United Kingdom }
\affiliation{London Centre for Nanotechnology, and Department of Electronic \& Electrical Engineering, University College London, Torrington Place, London, WC1E 7JE, United Kingdom }
\author{C.~Emary}
\affiliation{Department of Physics and Mathematics, University of Hull, Kingston-upon-Hull HU6 7RX, United Kingdom }
\author{P.~See}
\affiliation{National Physical Laboratory, Hampton Road, Teddington, Middlesex TW11 0LW, United Kingdom }
\author{J.~P.~Griffiths}
\affiliation{Cavendish Laboratory, University of Cambridge, J. J. Thomson Avenue, Cambridge CB3 0HE, United Kingdom }
\author{G.~A.~C.~Jones}
\affiliation{Cavendish Laboratory, University of Cambridge, J. J. Thomson Avenue, Cambridge CB3 0HE, United Kingdom }
\author{I.~Farrer}
\altaffiliation[Current address: ]{Department of Electronic \& Electrical Engineering, University of Sheffield, Mappin Street, Sheffield S1 3JD, United Kingdom }
\affiliation{Cavendish Laboratory, University of Cambridge, J. J. Thomson Avenue, Cambridge CB3 0HE, United Kingdom }
\author{D.~A.~Ritchie}
\affiliation{Cavendish Laboratory, University of Cambridge, J. J. Thomson Avenue, Cambridge CB3 0HE, United Kingdom }
\author{M.~Pepper}
\affiliation{London Centre for Nanotechnology, and Department of Electronic \& Electrical Engineering, University College London, Torrington Place, London, WC1E 7JE, United Kingdom }
\author{T.~J.~B.~M.~Janssen}
\affiliation{National Physical Laboratory, Hampton Road, Teddington, Middlesex TW11 0LW, United Kingdom }

\begin{abstract}
We report time-of-flight measurements on electrons travelling in quantum-Hall edge states. Hot-electron wave packets are emitted one per cycle into edge states formed along a depleted sample boundary. The electron arrival time is detected by driving a detector barrier with a square wave that acts as a shutter. By adding an extra path using a deflection barrier, we measure a delay in the arrival time, from which the edge-state velocity $v$ is deduced. We find that $v$ follows $1/B$ dependence, in good agreement with the $\vec{E} \times \vec{B}$ drift. The edge potential is estimated from the energy-dependence of $v$ using a harmonic approximation.
\end{abstract}

\date{\today}

\maketitle

Electronic analogues of fundamental photonic quantum-optics experiments, so-called ``electron quantum optics'', can be performed using the beams of single-electron wave packets. The demonstration of entanglement and multi-particle interference with such wave packets would set the stage for quantum-technology applications such as quantum information processing \cite{Bennett}. Various theoretical proposals \cite{Samuelsson,OlKhovskaya,Splettstoesser,Moskalets,Haack,Battista} and experimental realisations \cite{Ji,Henny,Feve,Bocquillon,Bocquillon2,Fletcher,Bocquillon3,Ubbelohde,Waldie,Freulon} employ quantum-Hall edge states \cite{Halperin} as electron waveguides. The group velocity and dispersion relation of edge states are important parameters for understanding and controlling electron wave-packet propagation. For edge-magnetoplasmons, the velocity can be deduced by time-of-flight measurements with fast gate pulses \cite{Ashoori,Zhitenev,Kamata,Kumada}. This is difficult with electron wave packets because gate pulses would also affect the background Fermi sea, and the experiments in the past \cite{McClure,Bocquillon3} use other transport data to estimate the electron velocity. Furthermore, electron-electron interactions can cause the formation of multiple collective modes travelling at different velocities, leading to fast decoherence \cite{Bocquillon3,Ferraro,Freulon}. In order to perform direct measurements of the bare group velocity of electron wave packets by time-resolved methods, we need a robust edge-state waveguide system where the interactions between the transmitted electrons and other electrons in the background can be suppressed.


\begin{figure}

\epsfxsize=8.5cm

\centerline{\epsffile{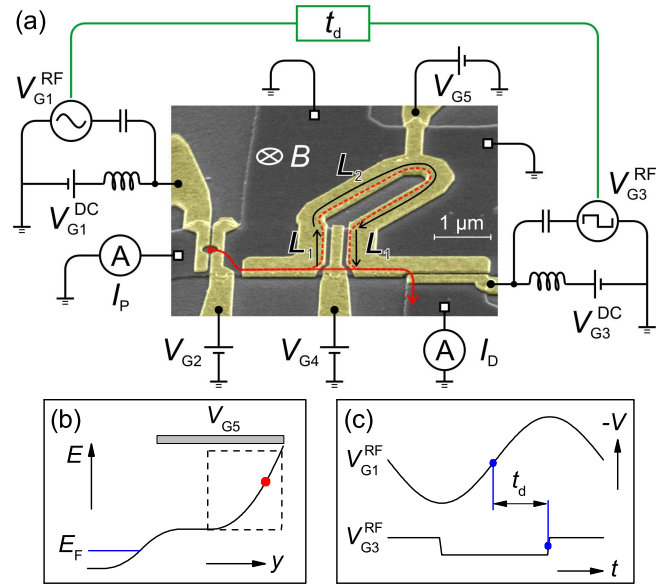}}

\vspace{-2mm}

\caption{(a) A scanning electron micrograph of a device and schematics of the measurement circuit used for time-of-flight measurements. (b) Schematics of the lowest Landau Level at the sample edge. The region under the depletion gate G5 is depleted, and hence the bottom of the Landau level is lifted above the Fermi energy $E_{\rm F}$. The electron wave packets travel at high energy states indicated by a red dot. The dashed box represents a region where a harmonic approximation is used to deduce the edge potential profile shown in Fig.~\ref{Fig3}(d). (c) Sine wave rf signal $V_{\rm G1}^{\rm RF}$ applied to the pump entrance gate G1 and square wave rf signal $V_{\rm G3}^{\rm RF}$ applied to the detector gate G3. Their relative phase delay $t_{\rm d}$ is controlled at a picosecond resolution using the internal skew control of the arbitrary waveform generator.}

\vspace{-5mm}

\label{Fig1}

\end{figure}

In this Letter, we demonstrate an experimental method for probing the bare edge-state velocity of electrons travelling in a depleted edge of a two-dimensional system. The electrons are emitted from a tunable-barrier single-electron pump \cite{Blumenthal,Kaestner1,Kaestner2} as hot single-electron wave packets ($\sim 100$~meV above the Fermi energy) \cite{Leicht,Fletcher}. These electrons are injected into an edge where the background two-dimensional electron gas (2DEG) is depleted to avoid the influence of electron-electron interactions. The arrival time of these wave packets is detected by an energy-selective detector barrier with a picosecond resolution \cite{Fletcher,Waldie}. The travel length between the source and detector is switched by a deflection barrier. The time of flight of the extra path is measured as a delay in the arrival time at the detector \cite{Deflector}. The edge-state velocity is calculated from the length of the extra path and the time of flight. We find that the edge-state velocity is inversely proportional to the magnetic field applied perpendicular to the plane of the 2DEG in a good agreement with the $\vec{E} \times \vec{B}$ drift velocity, where $\vec{E}$ is the electric field and $\vec{B}$ is the magnetic field.  We probe the dispersion of the edge states by controlling the electron emission energy. From the energy dependence of the velocity, we deduce the edge potential profile and obtain the information on the spatial positions of the edge states.

The measurements presented in this work are performed on two samples, Sample A and Sample B, with slightly different device parameters. Figure~\ref{Fig1}(a) shows a scanning electron micrograph of a device with the same gate design as Sample B. Both samples are made from GaAs/AlGaAs heterostructures with a 2DEG 90 nm below the surface, but the 2DEG carrier density is slightly different ($1.8 \times 10^{15}$~m$^{-2}$ for Sample A and $1.6 \times 10^{15}$~m$^{-2}$ for Sample B). The active part of the device is defined by shallow chemical etching and Ti/Au metal deposition using electron-beam lithography. The device comprises 5 surface gates: the pump entrance gate (G1), pump exit gate (G2), detector gate (G3), deflection gate (G4), and depletion gate (G5). $L_{1}$ is the path length along the deflection gate and is the same for both samples (1.5~$\mu$m). $L_{2}$ is the path length along the loop section defined by shallow etching, and is 2~$\mu$m for Sample A and 5~$\mu$m for Sample B. The measurements are performed in a cryostat with the base temperature at 300 mK.

Figure~\ref{Fig1}(a) also shows the measurement circuit. The rf sine signal $V_{\rm G1}^{\rm RF}$ (with a peak-to-peak amplitude $\sim 1$~V) applied to G1 pumps electrons over the barrier formed by the dc voltage $V_{\rm G2}$ applied on G2 \cite{Kaestner1}. The rf signal is repeated periodically at a frequency $f = 240$~MHz, producing the pump current $I_{\rm P}$. When the device pumps exactly one electron per cycle, $I_{\rm P} = ef \approx 38$~pA, where $e$ is the elementary charge. In a magnetic field $B$ applied perpendicular to the plane of the 2DEG [in the direction indicated in Fig.~\ref{Fig1}(a)], the electrons emitted from the pump follow the sample boundary and enter the region where the background 2DEG is depleted by the negative voltage $V_{\rm G5}$ on G5 (we apply $-0.45$~V for Sample A and $-0.3$~V for Sample B, well in excess of the typical depletion voltage of $\sim -0.2$~V). The bottom of the lowest Landau level is raised above the Fermi energy $E_{\rm F}$ but is kept lower than the electron emission energy as shown in Fig.~\ref{Fig1}(b). The electrons travel along the edge approximately 500~nm (roughly equal to the extent that G5 covers from the edge defined by shallow etching) away from the nearest 2DEG [as indicated by the red dot in Fig.~\ref{Fig1}(b)].  

Depending on the voltage $V_{\rm G4}$ applied to the deflection gate G4, the electron wave packets reach the detector (G3) either through the shorter route [solid red line in Fig.~\ref{Fig1}(a)] or the longer route (dashed line). In both cases, the majority of electrons reach the detector without measurable energy loss for the magnetic field considered here.  Electrons that lose energy through LO-phonon emission \cite{Fletcher,LOphonon} are reflected by the detector barrier and do not contribute to the detector current.  The longer route adds an extra length $2L_{1}+L_{2}$ to the electron path, causing a delay in the arrival time at the detector. The electron arrival time at the detector is detected using a time-dependent signal on the detector \cite{Fletcher,Waldie}. A square wave $V_{\rm G3}^{\rm RF}$ with a peak-to-peak amplitude of $\sim 20$~mV is applied to G3 \cite{Waldie}. The detector current $I_{\rm D}$ is monitored as $V_{\rm G3}^{\rm DC}$ is swept and the relative delay time $t_{\rm d}$ between $V_{\rm G1}^{\rm RF}$ and $V_{\rm G3}^{\rm RF}$ is varied with a picosecond resolution [see Fig.~\ref{Fig1}(c)] \cite{Waldie}.


\begin{figure}

\epsfxsize=8.5cm

\centerline{\epsffile{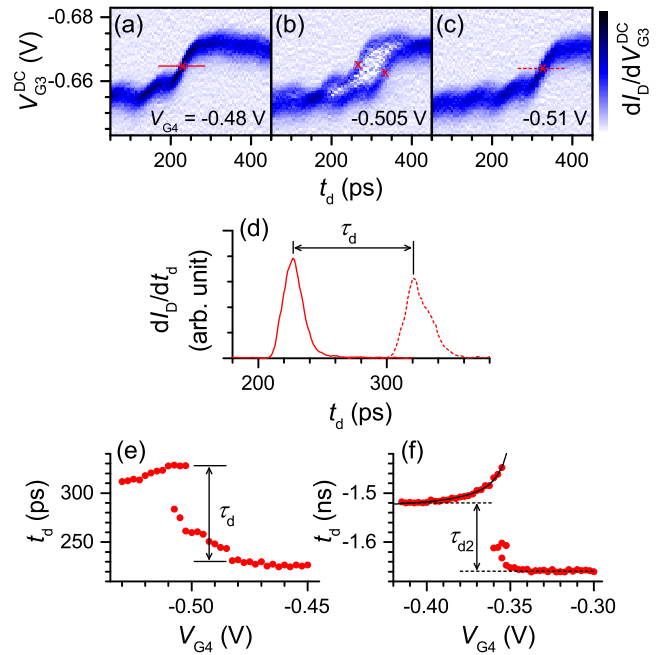}}

\vspace{-2mm}

\caption{ (a)-(c) $dI_{\rm D}/dV_{\rm G3}^{\rm DC}$ plotted in colour scale as a function of $V_{\rm G3}^{\rm DC}$ and $t_{\rm d}$ for three values of $V_{\rm G4}$. Red crosses are placed at the centre of the transition in the detector threshold, which indicates the peak in the electron arrival time. The data are taken from Sample A. (d) $dI_{\rm D}/dt_{\rm d}$ plotted as a function of $t_{\rm d}$ for two values of $V_{\rm G4}$: -0.48~V (solid line) and -0.51~V (dashed line). These peaks represent the arrival-time distributions, and the time difference $\tau_{\rm d}$ between them represents the time of flight.  (e) Peak arrival time plotted as a function of $V_{\rm G4}$. Taken from Sample A. (f) Peak arrival time taken from Sample B. The solid line is fit to $-1/V_{\rm G4}$.
}

\vspace{-5mm}

\label{Fig2}

\end{figure}

Figures~\ref{Fig2}(a)-(c) show the behaviour of the detector current for three values of $V_{\rm G4}$ taken at $B = 14$~T with Sample A. Here, $dI_{\rm D}/dV_{\rm G3}^{\rm DC}$ is plotted in colour scale as a function of $V_{\rm G3}^{\rm DC}$ and $t_{\rm d}$. The pump current is set at the quantised value for one electron emission per cycle (i.e.\ $I_{\rm P} \approx ef$) with $V_{\rm G1}^{\rm DC} = -0.46$~V and $V_{\rm G2} = -0.56$~V. When the detector barrier is sufficiently low (i.e.\ $V_{\rm G3}^{\rm DC}$ is less negative), all emitted electrons that do not suffer energy loss during the travel from the pump enter the detector contact and contribute to $I_{\rm D}$. Therefore, $I_{\rm D} \approx I_{\rm P}$ as the LO phonon emission is negligble at $B = 14$~T in these samples. When the detector barrier is sufficiently high, all electrons are blocked, and $I_{\rm D} = 0$.  When the detector barrier is matched to the energy of incoming electrons, a peak in $dI_{\rm D}/dV_{\rm G3}^{\rm DC}$ appears \cite{Fletcher,Waldie}. The peak position (or the detector threshold) in $V_{\rm G3}^{\rm DC}$ depends on $t_{\rm d}$ because a square wave is applied to the detector gate and the sum $V_{\rm G3}^{\rm DC}+V_{\rm G3}^{\rm RF}$ determines the detector barrier height. When $t_{\rm d}$ is small (large), electrons arrive when the square wave is negative (positive), and hence it shifts the detector threshold to more positive (negative) in $V_{\rm G3}^{\rm DC}$. The transition of the detector threshold in $V_{\rm G3}^{\rm DC}$ from more positive to more negative occurs at $t_{\rm d}$ where the square-wave transition coincides with the electron arrival at the detector. As $V_{\rm G4}$ is made more negative, the detector transition shows splitting [Fig.~\ref{Fig2}(b)] and finally settles to larger $t_{\rm d}$ [Figs.~\ref{Fig2}(c)]. The splitting happens as G4 splits the wave packets into the shorter and longer routes, and hence two sets of electron wave packets arrive at the detector with a time delay. The shift of the detector transition to larger $t_{\rm d}$ occurs because the longer route causes a delay in the arrival time.

In Fig.~\ref{Fig2}(d), $dI_{\rm D}/dt_{\rm d}$ is plotted as $t_{\rm d}$ is swept through the centre point of the detector transition marked by red crosses for the cases of $V_{\rm G4} = -0.48$~V (solid line) in Fig.~\ref{Fig2}(a) and -0.51~V (dashed line) in Fig.~\ref{Fig2}(c). These two curves represent the arrival-time distributions for the shorter and longer routes, and hence the time difference $\tau_{\rm d}$ between the two peaks is the time of flight of the extra path ($2L_{1}+L_{2}$) taken by the longer route. The edge-state velocity in the extra path can be calculated as $v = (2L_{1}+L_{2})/\tau_{\rm d}$. In this example, $v = 5~\mu{\rm m}/95~{\rm ps} = 5.3 \times 10^4$~m/s. 

The uncertainty in the velocity measurement arises from the uncertainties in $2L_{1}+L_{2}$ and $\tau_{\rm d}$. The value of $2L_{1}+L_{2}$ is likely to be accurate only to $\pm 10$\% as we can only estimate it from the device geometry. This gives the same systematic error to all velocity estimates within the same sample, and hence it does not affect the discussions in the later sections qualitatively. The uncertainty in $\tau_{\rm d}$ is more problematic.  This is because the arrival time does not just switch between two values as the edge-state path is switched. As plotted in Fig.~\ref{Fig2}(e), the arrival time initially changes slowly towards larger $t_{\rm d}$ as $V_{\rm G4}$ is made more negative.  Then it starts to move through a series of small steps, seemingly in a random manner, until it makes a final large step. After that, the arrival time moves gradually back to smaller $t_{\rm d}$.  This behaviour can be interpreted as follows.  The series of changes for $-0.51 < V_{\rm G4} < -0.48$~V occur as the path length and the velocity of the edge state under G4 is altered in a complicated manner due to disorder potential. This lasts until the edge state is finally pushed out of the region under G4 at $V_{\rm G4} \sim -0.51$~V, and is switched to the longer route. Then the arrival time continues to change as $V_{\rm G4}$ is made more negative, because the velocity along G4 keeps increasing as the potential profile along G4 is made steeper (due to the $\vec{E} \times \vec{B}$ drift as discussed later).  For the measurements with Sample A, we take $\tau_{\rm d}$ to be the difference in arrival time before and after rapid changes as indicated in Fig.~\ref{Fig2}(e).  A typical uncertainty in $\tau_{\rm d}$ estimate by this method is $\pm 5$~ps.

A more rigorous velocity estimate can be introduced by excluding the contribution from the electron paths along G4. Figure~\ref{Fig2}(f) shows the time-of-flight data taken from Sample B plotted in the same manner to Fig.~\ref{Fig2}(e). With Sample B, the arrival time changes more rapidly as $V_{\rm G4}$ is made more negative after the electron path is switched to the longer route. As the case with Sample A, this is considered to result from a rapid change in the velocity along G4, and is the main source of the uncertainty in velocity estimates. In order to reduce the uncertainty, we break up the time of flight into two parts, $\tau_{\rm d1}$ along G4 (length $2L_{\rm 1}$) and $\tau_{\rm d2}$ along the loop (length $L_{\rm 2}$), i.e. $\tau_{\rm d} = \tau_{\rm d1} + \tau_{\rm d2} = 2L_{\rm 1}/v_{\rm 1} + L_{\rm 2}/v_{\rm 2}$, where $v_{\rm 1(2)}$ is the velocity along the path $L_{\rm 1(2)}$. From this, one can see $\tau_{\rm d} \rightarrow \tau_{\rm d2} = L_{\rm 2}/v_{\rm 2}$ in the limit $v_{\rm 1}/v_{\rm 2} \gg 2L_{\rm 1}/L_{\rm 2}$. Once the electron path is deflected, $v_{\rm 1}$ increases as $V_{\rm G4}$ is made more negative, whereas $v_{\rm 2}$ is unaffected. Therefore, in the limit of large negative $V_{\rm G4}$, the time of flight settles to $\tau_{\rm d2}$, the time of flight around the loop section. It is not trivial to know exactly how $v_{\rm 1}$ changes with $V_{\rm G4}$, but a linear relation ($v_{\rm 1} \propto -V_{\rm G4}$) fits well to the experimental data [solid line in Fig.~\ref{Fig2}(f)]. As shown in Fig.~\ref{Fig2}(f), $\tau_{\rm d2}$ can be estimated as the difference between the saturated values of arrival time at the positive and negative ends of $V_{\rm G4}$. The velocity around the loop section is calculated as $v_{\rm 2} = L_{\rm 2}/\tau_{\rm d2}$. We find that the uncertainty is reduced approximately by a factor of three using this method. We note that we cannot apply this method to Sample A as $L_{\rm 2}$ is too small to observe the saturation in the arrival time at the negative end in $V_{\rm G4}$.


\begin{figure}

\epsfxsize=8.5cm

\centerline{\epsffile{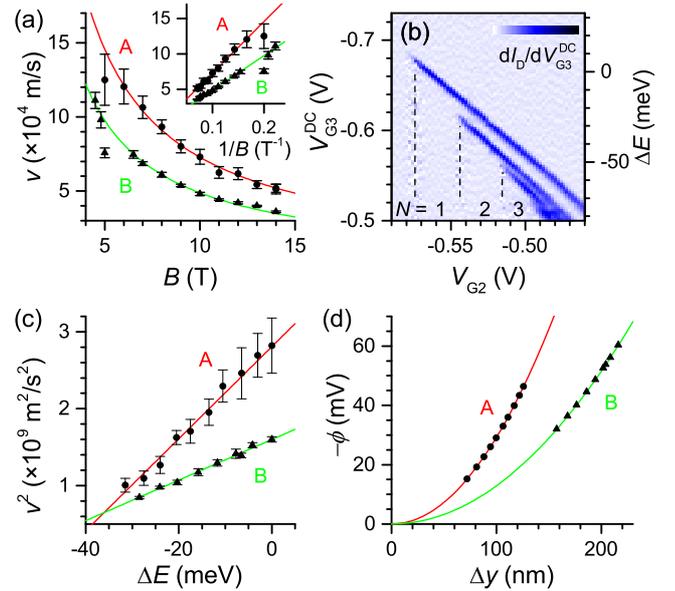}}

\vspace{-2mm}

\caption{(a) Edge-state velocity $v$ measured as a function of $B$. Circle (triangle) data points are taken with Sample A (B). The solid curves are fits to $1/B$. Inset: $v$ plotted against $1/B$. (b) Electron emission-energy spectrum measured at $B = 14$~T. (c) $v^2$ plotted as a function of relative emission energy $\Delta E$. Solid lines are fits to a linear relation. (d) Edge confinement potential $-\phi$ (solid lines) estimated from the velocity measurements. The spatial positions of the edge states corresponding to the velocity measurements are indicated by symbols.}

\vspace{-5mm}

\label{Fig3}

\end{figure}

Now, we investigate the magnetic-field and emission-energy dependence of the velocity to see if the studied edge-state system is consistent with a simple, interaction-free, quantum-Hall edge-state model. Figure~\ref{Fig3}(a) shows the magnetic field dependence of the measured edge-state velocity for both samples [$v$ (the velocity along the whole extra path, $2L_{1}+L_{2}$) is plotted for Sample A and $v_{\rm 2}$ (the velocity along the loop section only, $L_{2}$) for Sample B]. Clear $1/B$ dependence is observed for both samples down to $B = 5$~T. This is in good agreement with the $\vec{E} \times \vec{B}$ drift velocity, where $v = |\vec{E} \times \vec{B}|/B^2 \propto 1/B$, and $\vec{E}$ is the electric field due to the edge potential.  For Sample A, the data point at $B = 5$~T deviates from the $1/B$ dependence. This may be a sign of crossover from the skipping orbits to the $\vec{E} \times \vec{B}$ drift observed by McClure {\it et al} \cite{McClure}, but also this could be simply because of a large uncertainty due to a shorter time of flight.

In order to estimate the edge confinement potential and the spatial position of edge states, we consider the dispersion relation in a quasi-one-dimensional channel with a harmonic approximation \cite{Dispersion}. For the lowest branch of magneto-electric subband \cite{Datta},
\begin{equation}
E(k_{x}) = \epsilon_0 +\frac{1}{2}\hbar \Omega + \frac{\hbar^2 k_{x}^{2}}{2m^*}\frac{\omega_{y}^{2}}{\Omega^2},
\label{dispersion}
\end{equation} 
where $k_{x}$ is the wave number in the edge-state transport direction (in $x$ direction), $\epsilon_0$ is the lowest two-dimensional subband energy, $\Omega = \sqrt{\omega_{y}^{2}+\omega_{c}^{2}}$, $\hbar \omega_{y}$ is the transverse confinement energy (in $y$ direction), $\hbar \omega_{c}$ is the cyclotron energy, and $m^{*}$ is the electron effective mass. From the dispersion relation and the group velocity $v = 1/\hbar \cdot dE/dk$, one can deduce 
\begin{equation}
\frac{1}{2}m^{*}v^{2} \approx \frac{\omega_{y}^{2}}{\omega_{c}^{2}} \left( E-\epsilon_0 - \frac{1}{2}\hbar \omega_{c} \right),
\label{velocityrelation}
\end{equation}
in the limit of large magnetic field ($\omega_c \gg \omega_y $). From this, $\omega_{y}$ can be deduced by plotting $v^2$ against $E$.

The electron emission energy from our single-electron source can be tuned over a wide range \cite{Fletcher}. This can be used to probe the energy dependence of the edge-state velocity.  Figure~\ref{Fig3}(b) shows the emission energy spectrum measured as $V_{\rm G2}$ is varied with a static detector barrier ($V_{\rm G3}^{\rm RF} = 0$) with electrons travelling along the longer route ($V_{\rm G4} = -0.7$~V) at $B = 14$~T.  The conversion to relative emission energy $\Delta E$ [shown on the right vertical axis in Fig.~\ref{Fig3}(b) with the highest energy point used in this work set at zero] is made by calibrating $V_{\rm G3}^{\rm DC}$ against LO-phonon emission peaks \cite{Fletcher} (not visible in this particular dataset) and assuming the LO phonon energy of 36~meV \cite{Heiblum}. The electron emission energy decreases linearly as $V_{\rm G2}$ is made more positive. When $V_{\rm G2}$ is swept further positive, multiple ($N$) electrons are pumped per cycle. Even then, we can still resolve the emission energy of the last electron emitted, as the emission energies of other electrons are well separated at lower values and can be blocked by the detector.

Figure~\ref{Fig3}(c) plots $v^{2}$ measured as a function of relative emission energy $\Delta E$ at $B = 14$~T for both samples. As expected, they fit well to straight lines. From this fit, we deduce the edge-confinement energy $\hbar \omega_y = 2.7$~meV and 1.8~meV, and the bottom of the confinement potential at $\Delta E = -47$~meV and $-61$~meV, for Samples A and B, respectively. From these, we can reconstruct the edge-confinement potential $\phi = - m^{*}\omega_{y}^{2} y^2 / 2e$ as shown in Fig.~\ref{Fig3}(d). Here we set the potential at the bottom of the parabola as zero.  From each data point in Fig.~\ref{Fig3}(c), we can deduce the potential energy $-e\phi$ at the position of the guiding centre, averaged over the length of the path, by subtracting the kinetic energy $\frac{1}{2}m^{*}v^{2}$ from the total (relative) energy $\Delta E$. We can then visualise the spatial position of the edge states as plotted in Fig.~\ref{Fig3}(d).

In summary, we have shown the measurements of the time of flight of electron wave packets travelling through edge states. The electrons travel in the region where the background 2DEG is depleted and electron-electron interaction is minimised. We find that the electron velocity is in good agreement with the expected $\vec{E} \times \vec{B}$ drift. From the energy dependence, we deduce the edge confinement potential. Our technique provides a way of characterising the edge-state transport of single-electron wave packets with picosecond resolutions. The method that we have developed to transport electron wave packets in depleted edges could provide an ideal electron waveguide system where decoherence due to interactions can be avoided for electron quantum optics experiments.

\begin{acknowledgments}
We would like to thank Heung-Sun Sim and Sungguen Ryu for useful discussions. This research was supported by the UK Department for Business, Innovation and Skills, NPL's Strategic Research Programme, and the UK EPSRC.
\end{acknowledgments}


\begin{thebibliography}{}
\bibitem{Deflector}				One may argue that time-of-flight measurements could be performed by using two detector gates on the electron path, and detecting the difference in the arrival time between the two detectors. However, in this method, synchronising two detector signals with a picosecond resolution would be challenging. Our method circumvents such a timing issue by using a single detector.
\bibitem{LOphonon}				We find that the depletion gate can be used to suppress the energy relaxation due to LO-phonon emission \cite{Fletcher}. With a negative voltage on G5 enough to deplete the 2DEG underneath, LO-phonon scattering becomes negligible down to ($B \sim 7 - 8$~T). Even at lower fields where LO-phonon scattering is enhanced, we can still perform velocity measurements as an enough number of electrons arrive at the detector with the original emission energy. We believe that the suppression of LO-phonon emission is caused by the decreased overlap between the initial and final states due to shallower slope at the bottom of parabolic potential. The details of this effect will be published elsewhere \cite{Emary}.
\bibitem{Dispersion}			Although, in this approximation, the channel is confined by a harmonic potential whereas in our device the confinement potential is only on one side (the other edge is far away), it is still appropriate to use this approximation in high $B$ limit as the wave function becomes localised on one edge only.
\end{thebibliography}
\end{document}